# The Necessity for a Time Local Dimension in Systems with Time Varying Attractors

*K. Saermark*

Physics Laboratory I, The Technical University of Denmark, DK-2800 Lyngby, Denmark

*Y. Ashkenazy* [1], *J. Levitan* [1] [2] *and M. Lewkowicz* [1] [2]

(1) Department of Physics, Bar-Ilan University, 52900 Ramat-Gan, Israel

(2) The Research Institute, The College of Judea and Samaria, 44837 Ariel, Israel

**Abstract**

We show that a simple non-linear system of ordinary differential equations may possess a time varying attractor dimension. This indicates that it is infeasible to characterize EEG and MEG time series with a single time global dimension. We suggest another measure for the description of non-stationary attractors.

1. Introduction

In the study of the chaotic behavior of a dynamic system the calculation of the (correlation) dimension plays a central role. Briefly stated this quantity yields the minimum dimension of a space that can contain the trajectories generated by the system, and its calculation is a valuable first step in the study of the system considered. Further, for a given experimental system an examination of possible changes in the dimension under different operational conditions is of interest.

In recent years such calculations have been performed for example on electroencephalographic (EEG) and magnetoencephalographic (MEG) time-series recordings [1]-[4]. It has been conjectured that complex, aperiodic behavior of such time series it due to deterministic chaos [1]. Although a general consensus as to the clinical implications of a "dimension" does not seem to exist, there is some evidence for a change in this value in going from normal subjects to neurological patients (see, e.g., [2], [5]). In such calculations one has, most often, used the Grassberger-Procaccia algorithm, a procedure which may, however, be somewhat dubious to apply - as it requires extremely long time-series - to a system of such an enormous complexity as the human brain. Even within one and the same measurement one often finds an abrupt change in the complex pattern of the time series which indicates that it is impossible to ascribe one single value as representative for the (time) global dimension. If the above mentioned change of dimensionality can be taken as indicative of the existence of different attractors in brain activity, as reflected in EEG and MEG, then, clearly, there is a need for a description in terms of a time local dimensionality [6]. In the following we investigate the possibility of consistently defining and using such a temporal definition. To elucidate this point we have performed computer integrations of a set of equations consisting of a superposition of two well-known chaotic systems.

We shall in this study

1) discuss a non linear system of ordinary differential equations that bears a high degree of sensitivity to small changes in one of the coefficients,

2) discuss a varying time dependence of the solution of the underlying differential equations,

3) carry out a local analysis of the 'dimension'.

Referring to 1), one finds intervals for the values of the changeable coefficient where this sensitivity is much higher than for values outside these intervals. It turns out, that exactly in these intervals the system undergoes marked changes in its time dependence. We moreover remark that this time dependent behavior is connected with the phenomenon of intermittency or bursts of chaos [7-8]. We shall demonstrate that even though the data generator is not altered once the above mentioned coefficient is fixed, our system shows a time variation in behavior. This behavior questions the normal practice of attaching a certain dimension to the attractor of a time series and points to the necessity of defining different dimensions for smaller time sections of the series.

We shall present here a single model which possesses a time varying attractor dimension. Such an example has to the best of our knowledge not been exhibited before although it has been conjectured that non stationarity of the attractor obscures the idea of an attractor dimension [9-11]. We stress that the idea of attaching distinct values to different time stages of the (chaotic) scenario may turn out to be of importance in physiological systems where the relative increase/decrease of the dimension may reflect an actual change in the event related potentials. We emphasize that fixed coefficients as in this study do not account for such time varying generators.

## 2. The dependence on the parameters

The model considered here is given by the following equations

(1)
$$\frac{dx}{dt} = \alpha(-y-z) + \beta\sigma(y - y_0 - x + x_0)$$
$$\frac{dy}{dt} = \alpha(x + ay) + \beta((x_0 - x)(z - z_0) + r(x - x_0) + y_0 - y)$$
$$\frac{dz}{dt} = \alpha(b + z(x - c)) + \beta((x - x_0)(y - y_0) - B(z - z_0))$$

For $\alpha = 1$ and $\beta = 0$ this system reduces to the Roessler system and for $\alpha = 0$ and $\beta = 1$ it reduces to the Lorenz system. For simplicity this study will be confined to values on the unit circle $\alpha^2 + \beta^2 = 1$. We keep the following parameters fixed: a, b, c, r, B, $x_0$, $y_0$, $z_0$ and $\sigma$. The specific values used are given in Table 1. Only the coefficient $\alpha$ (and thus $\beta$) is varied.

In this section we will display the trajectory of system (1) for the various values of the coefficient $\alpha$.

For $\alpha$ close to zero, one should expect a trajectory not much different from that of the pure Lorenz system. Indeed, the Lorenz-like attractor in fig. 1a was obtained with an $\alpha$ value of 0.035.

A slight change in the value of the coefficient $\alpha$, even only up to 0.04, changes the system's trajectory abruptly. Instead of oscillating in a Lorenz-like manner for a long time (more than 6000 time units, where $\alpha$=0.035; all the integrations were performed with a time step dt = 0.001) the trajectory for $\alpha$=0.04 in fig. 1b approaches a fixed point and starts immeaditely to spiral towards this fixed point.

This type of convergence of the trajectory to a fixed point is characteristic for the system's solution for all $\alpha$ values up to approximately 0.965 (see fig. 2a). Above this value until ca. 0.97 the trajectory consists of a limit cycle around a different fixed point of the system. Convergence towards this fixed point was not observed for 3200 time units. Increasing the $\alpha$ value further, one obtains a trajectory advancing more and more directly to this fixed point. (fig 2b for $\alpha=0.995$) Further increasing $\alpha$ unveils a new scenario of a stronger and stronger spiraling towards a yet different fixed point. (fig 2c and 2d) until $\alpha$ equals approximately 0.999995 where there is a singular transition to an outwards spiraling trajectory (fig 2e). Finally for $\alpha=0.999999$ the behavior evolves into a Roessler-like trajectory (fig 2f), which eventually leads to the well-known Roessler attractor for $\alpha=1.0$.

3. The time dependence

We shall now show that the dimension of the system's attractor exhibits a high degree of time dependence for a fixed value of the coefficient $\alpha$ if this value is chosen near the above mentioned transition from a Lorenz-like behavior to the convergence to a fixed point. Specifically we applied the value of $\alpha = 0.037005$, but similar behavior was found for all values of $\alpha$ belonging to the interval $0.0370 < \alpha < 0.0375$.

In figure 3 we display the trajectory in the x-z plane for different time intervals. For the first 600 time units (60000 data points) the trajectory is a limit cycle. Towards the very end of this time interval it starts to spiral outward (fig 3a).

In the interval from 600 to 3154 time units (255400 data points) a Lorenz-like behavior is noticeable (fig. 3b). This behavior is however limited in time and followed by an approach to a fixed point which can be seen from figure 3c, where the time interval from 3150 to 4000 time units ( 85000 data points ) is covered.

The period between the approximate limit cycle and the run into the static state resembles the well-known phenomena of intermittency. From fig 4a, showing the oscillations of the x-component, it is seen that the pronounced peaks coincide with the transitions from fig 3a to 3b and from fig 3b to fig 3c. The first burst of intermittent behavior is hence observed at the time where the system makes a transition from a more stable state to a more complex type of trajectory. After the transition to the more complex mode, intermittent oscillations continue for a considerable time, and then die out with the transition from the complex Lorenz like attractor to the simple approach to the fixed point.

We computed the dimensions for the different time sections shown in fig. 3. By applying a Grassberger-Procaccia type algorithm for the Generalized Correlation Dimension one finds a dimension for the first 600 time units (i.e. 60000 data points, the trajectory shown in fig. 3a) of 1.35. For the intermediate part (fig 3b) a dimension of 1.78 is obtained, which is in good accordance with the Lorenz-like trajectory taking into consideration that the Lorenz attractor has a dimension of 2.07 [26]. Finally for the very long time behavior (fig 3c) exhibiting the convergence to one of the limiting points a dimension is not defined. (For an example of and details about the method of the calculation see [12,14,17,18-22].)

All the obtained values by the Grassberger-Procaccia type algorithm were supported by the almost identical results obtained by using the Generalized Box-Counting Method [12].

On the basis of the above results which differ from one time interval to another we are led to the conclusion that an application of the standard methods for the calculation of the correlation dimension for the entire time series is meaningless.

This problem might be avoided by defining and calculating a (time) local dimension [11]. The present model is then best characterized by a dimension of 1.35 for the first 600 time units, whereas for the interval between 600 and 3154 time units a quite different value, 1.78, is found.. For the third time interval, wherein the trajectory approaches a fixed point, a correlation dimension is not defined. We have thus demonstrated that for a simple system in which all the parameters are being kept fixed, the dimension estimation varies with the time interval chosen

4. Discussion

The value, 1.78, which was obtained for the dimension, for the intermediate part of the scenario shown in fig 3a - 3c, is in good accordance with the value of ca 2.07 which is the dimension of the pure chaotic Lorenz attractor [16]. Chaotic behavior is characterized by a non integer dimension larger than 2 [27]. Taking into consideration , that our system does not exhibit pure chaos, but the less complex behavior of intermittency, one would expect a value slightly smaller than 2 for the Lorenz attractor like intermediate part of the trajectory. A number even smaller is anticipated for the initial part, which is clearly less complex.

All the computations of the correlation dimensions were performed with a number of data points exceeding by far what is generally assumed necessary in order to obtain a credible estimate. It is a well-know claim, that a dimension of ~2 necessitates a number of data points $\geq 42^2$ [13]-[15].

The use of the correlation dimension as a diagnostic tool for the detection of abnormal states in some physiological systems (as for example the brain) has been questioned by many authors [4,11,23] due to the non stationarity of the signal measured. The difficulties encountered are easily recognized from the big variance for the dimensions obtained by different researchers while measuring under similar conditions [9,10], so that the huge observed discrepancies in the results seem not justified. The present work does indeed underline the lack of sense in attaching one dimension value to an entire physiological time signal, which does not remain stationary over a long enough interval.

It may very well be, that much information can be obtained from the relative shift in the values obtained from the different time segments. It has already been suggested, that the dimensional shift following a stimulus does not change to a new stable state but actually undergoes a variety of rapid non-stationary changes [9]. Studies in event related potentials seem to confirm such a conjecture .

Following the above observations we suggest as a potential measure the information entropy [22,24], which may characterize the dynamics of non-stationary time series. In fig 4b we have displayed the time evolution of the information entropy $I_1$[1]. We choose this measure due to its sensitivity to the local structure of the phase space [28]. It is seen that a sharp increase in the information entropy takes place exactly for the times where intermittency can be observed. It seems therefore reasonable that this entropy can be used both to describe the onset of each burst of chaos and also indicates their relative intensity. The varying value of the information entropy underlines once more the need for a time local notion in the description of the these types of systems.

The present computer simulation study was motivated by an attempt to understand why attractor dimension calculations for example from EEG and MEG time series lead to so many values as found in the literature ( in general by applying the Grassberger-Procaccia algorithm). In recent years much work has been performed in order to identify attractors for various states of activities in biological systems, in particular brain activity. The question is whether it is possible to verify the existence of chaotic or strange attractors in brain activity measured by EEG and MEG recordings as suggested by the pioneering work of Babloyantz et al [1]. The results are, as mentioned, not unambiguous, but there does seem to be some evidence for lower correlation dimensions for some neurological diseases like epilepsy (e.g. [4]) or Schizophrenia (e.g.[5[) relative to those for healthy subjects.

---

[1]The entropy shown in fig 4b was obtained repeatedly for consecutive time segments of the time series x(t), each of 50 time units (5000 data points) with an overlap of half a time segment. Note, that in each time segment the data points were normalized to fall in the range 0-1.

It appears reasonable to assume that a non stationary system of high complexity, such as the brain, cannot be described by a single attractor, but that the underlying system of equations reflects a superposition of two or more systems each one describing its physiological generator, with, in addition, time varying parameters.

The system considered here has constant parameters and has been chosen vastly simpler than a brain system. In spite of this, as illustrated in fig 4, there are solutions for certain choices of parameters, where the resulting time series exhibits very fluctuating behavior, with a period of rather regular oscillations followed by a burst of chaotic like activity which fades off after a while. We therefore conclude that in general it does not make sense to calculate global dimensions for such systems, let alone to introduce minor corrections to the Grassberger-Procaccia algorithm [25]. Instead it appears necessary to use some kind of temporal measure for the correlation dimension of a non stationary time series or to use the information entropy as a measure as it is suggested in the present study.

| | |
|---|---|
| a | 0.15 |
| b | 0.20 |
| c | 10.00 |
| r | 45.92 |
| B | 4.00 |
| $\sigma$ | 16.00 |
| $x_0$ | -0.003001 |
| $y_0$ | 0.02001 |
| $z_0$ | -0.02001 |

Table 1: The parameters for the system (1)

**Figure Captions**

Fig 1. Trajectories in the x-z plane

(a) $\alpha=0.035$, t=164 (16384 time steps).

(b) $\alpha=0.040$, t=41 (4096 times steps).

Fig 2. Different types of trajectories in the x-y plane for various values of $\alpha$.

(a) $\alpha = 0.965$, t=41 (4096 time steps)

(b) $\alpha = 0.995$, t=41 (4096 time steps)

(c) $\alpha = 0.999$, t=41 (4096 time steps)

(d) $\alpha = 0.9995$, t=41 (4096 time steps)

(e) $\alpha = 0.999995$, t=54 (1080 time steps)

(f) $\alpha = 0.999999$, t=1638 (16384 time steps)

Fig 3. Different time segments of the trajectoery for $\alpha$ fixed. $\alpha=0.037005$

(a) $[t_1, t_2]$: $t_1=0$ and $t_2=600$ time units (60000 time steps)

(b) $[t_1, t_2]$: $t_1=600$ and $t_2=3154$ time units (255400 time steps)

(c) $[t_1, t_2]$: $t_1=3150$ and $t_2=4000$ time units (85000 time steps).

Fig 4. The time evolution for $\alpha = 0.037005$ of the

a) x coordinate

b) information entropy $I_1$ for increasing embedding dimensions from n=2 (the lower curve) to n=7 (the upper curve).

Note that the oscillatory behavior in fig 4a and 4b covers the same time interval as fig. 3b displaying the Lorenz-like attractor.

Fig. 1a

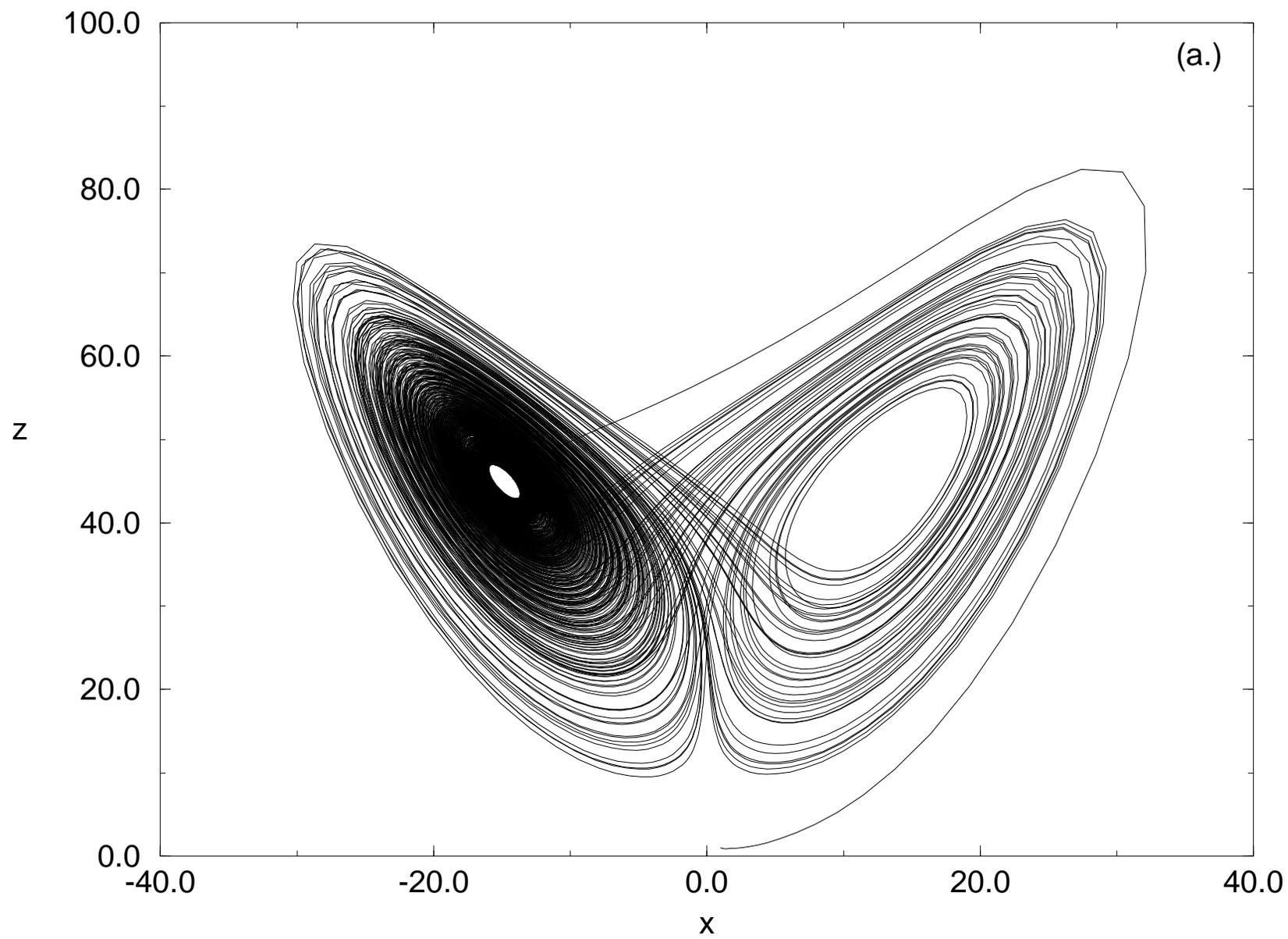

# Fig. 1b

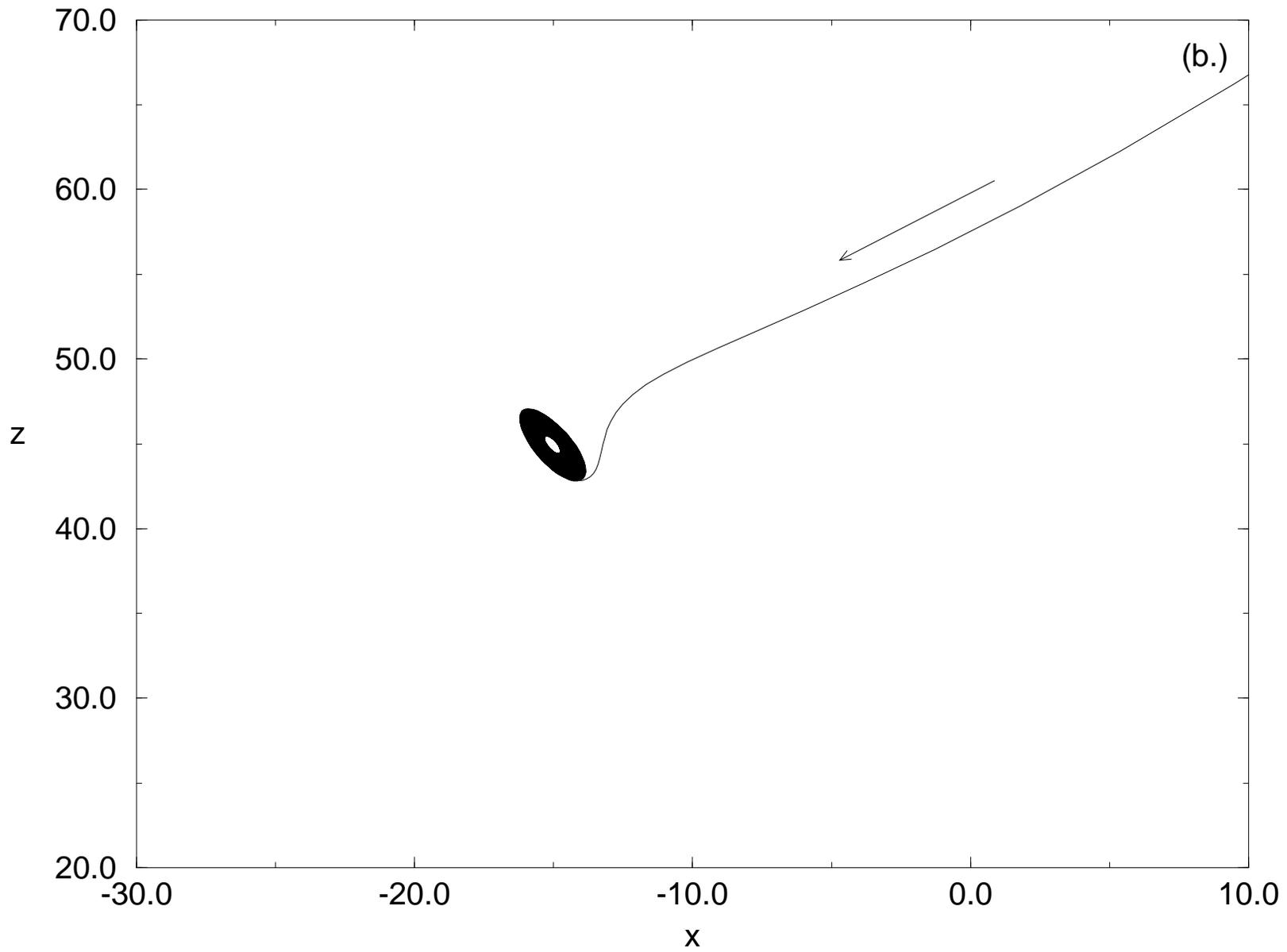

# Fig. 2a

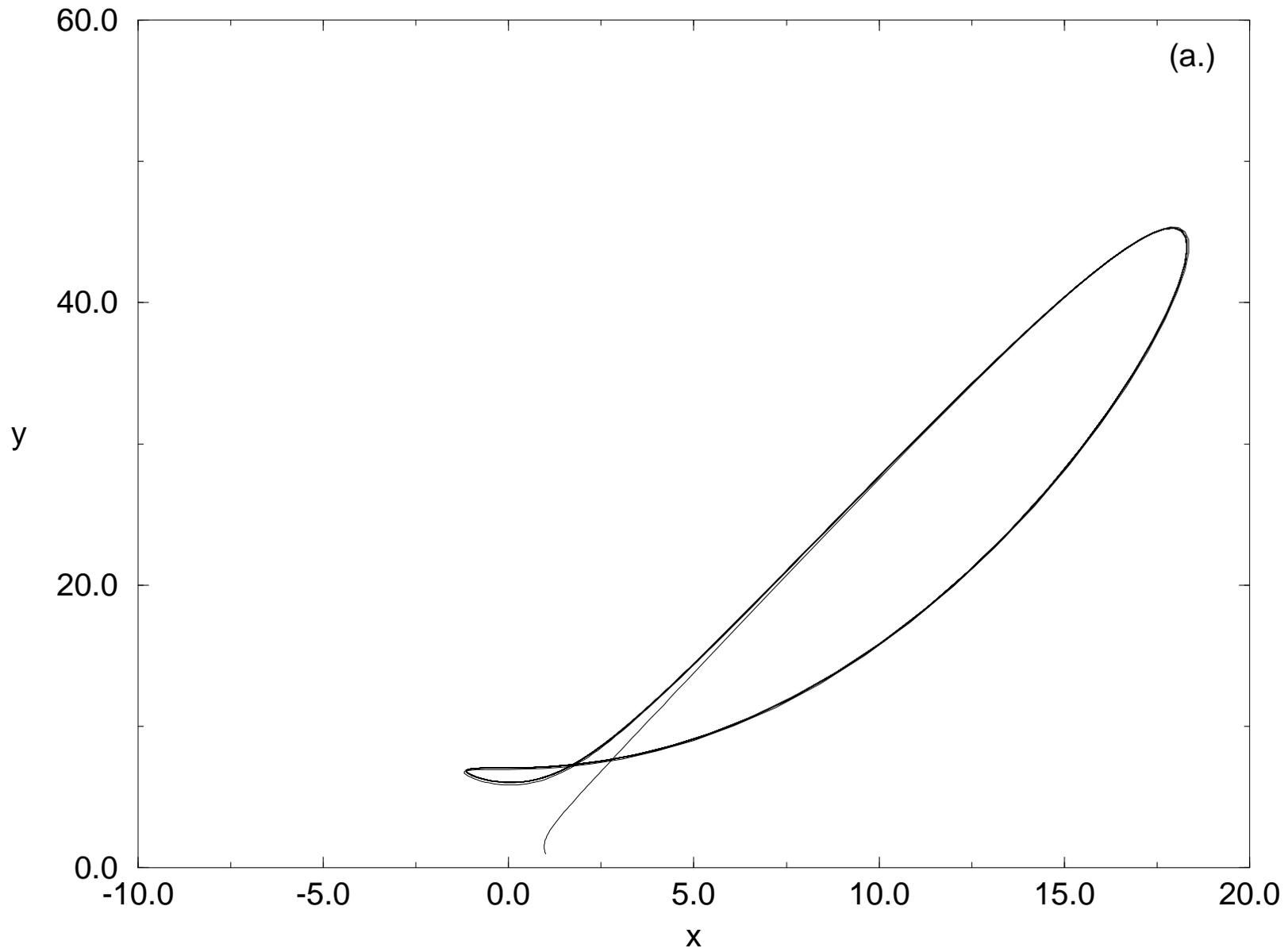

# Fig. 2b

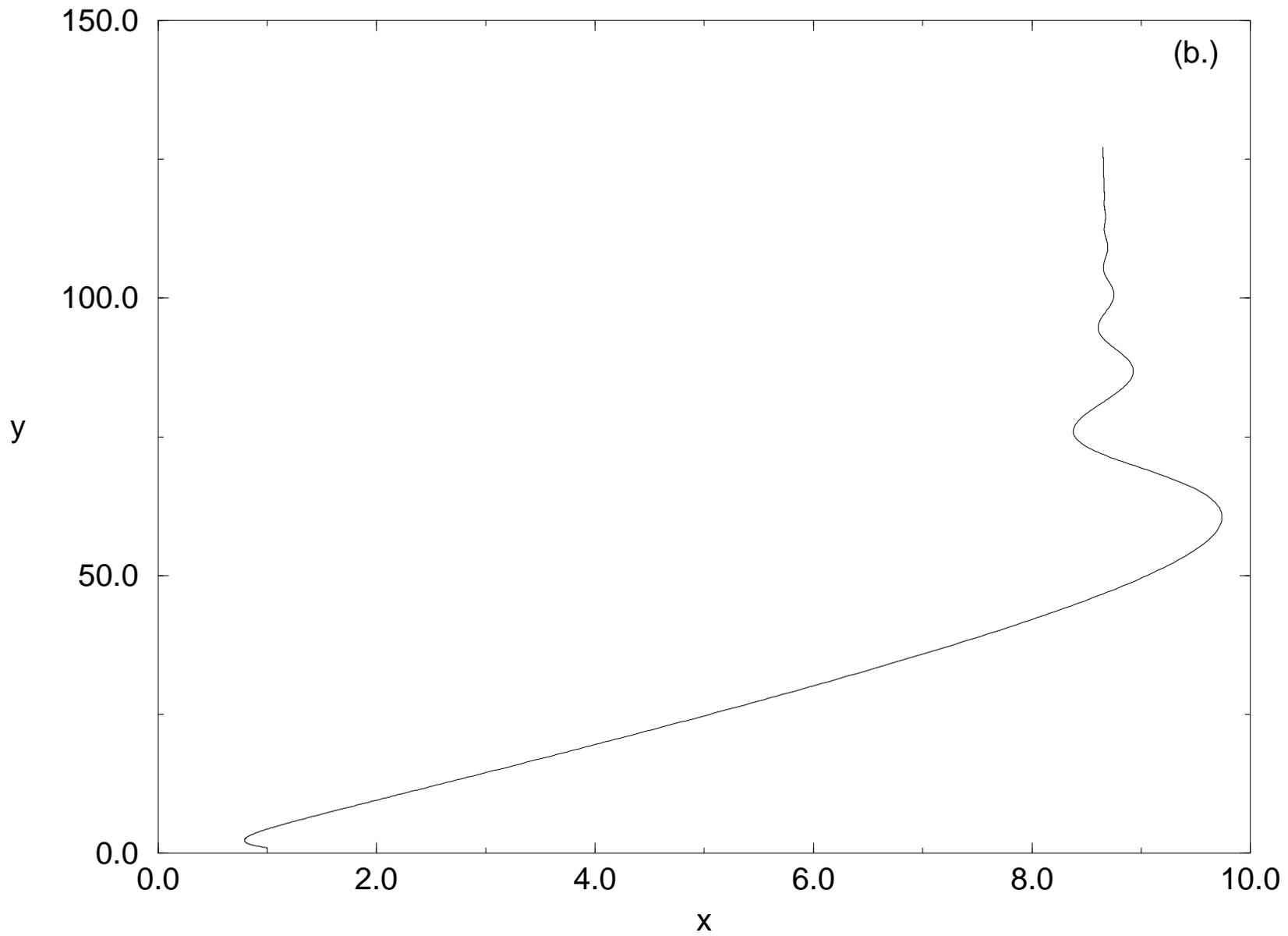

Fig. 2c

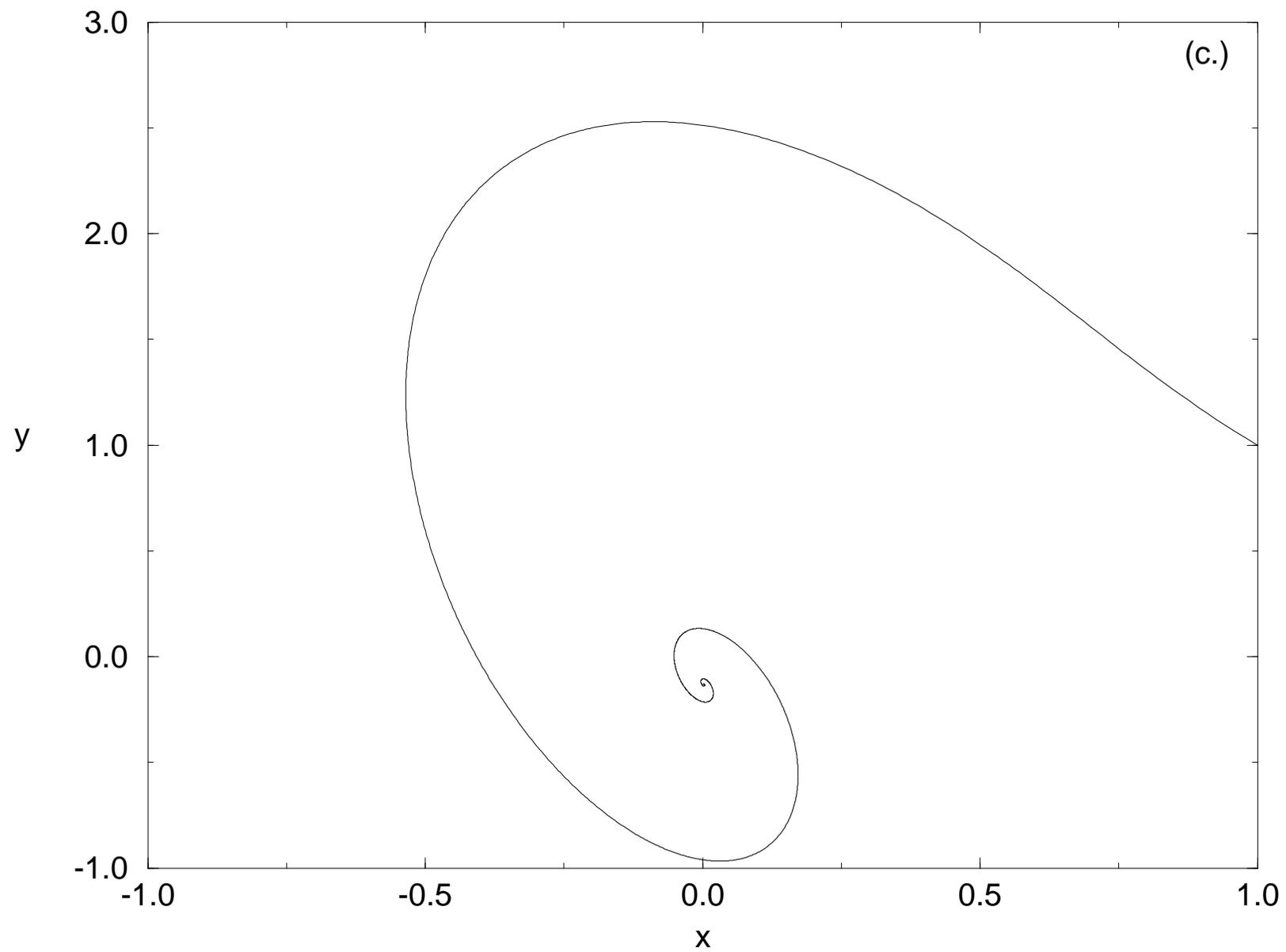

# Fig. 2d

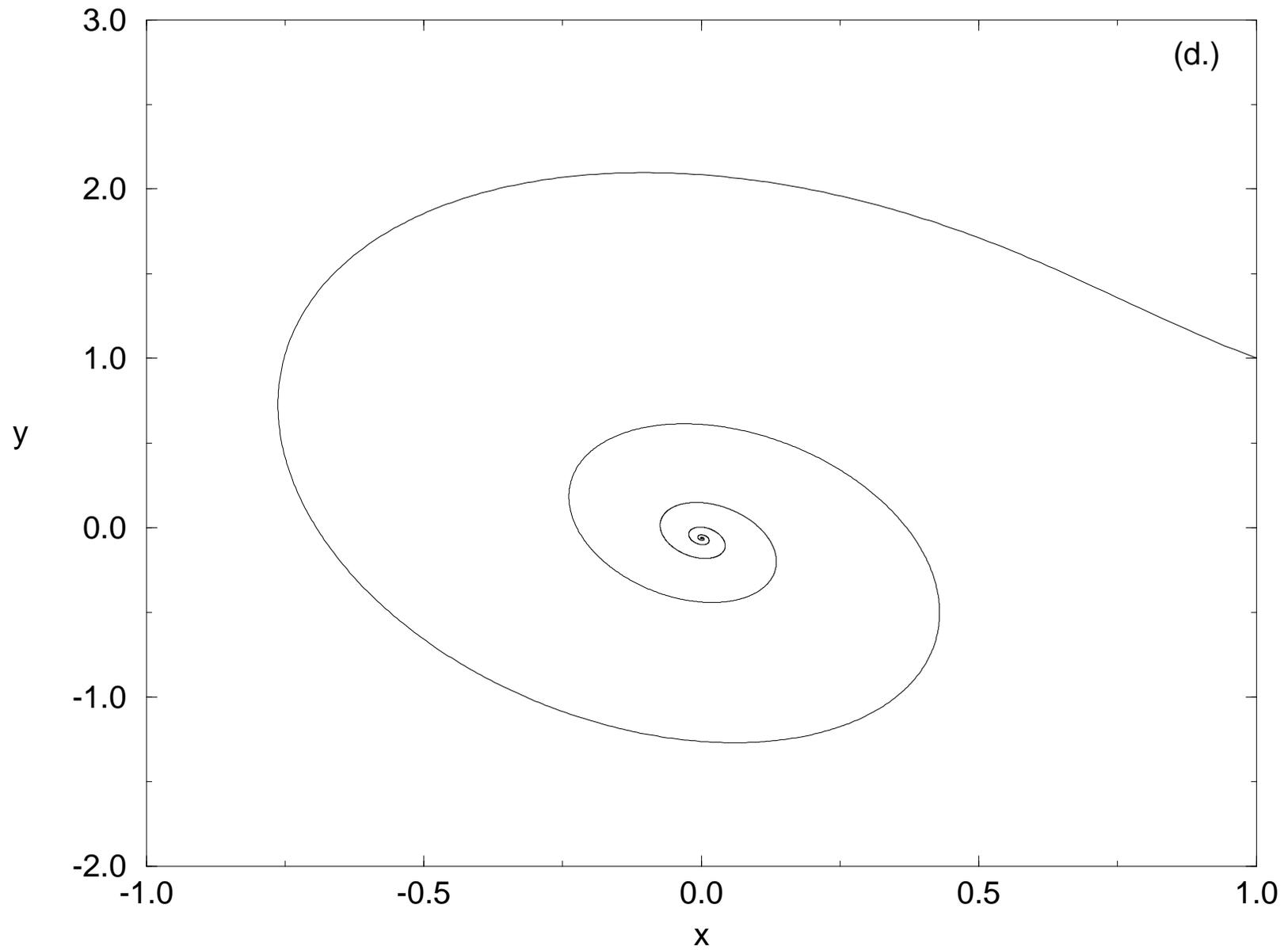

Fig. 2e

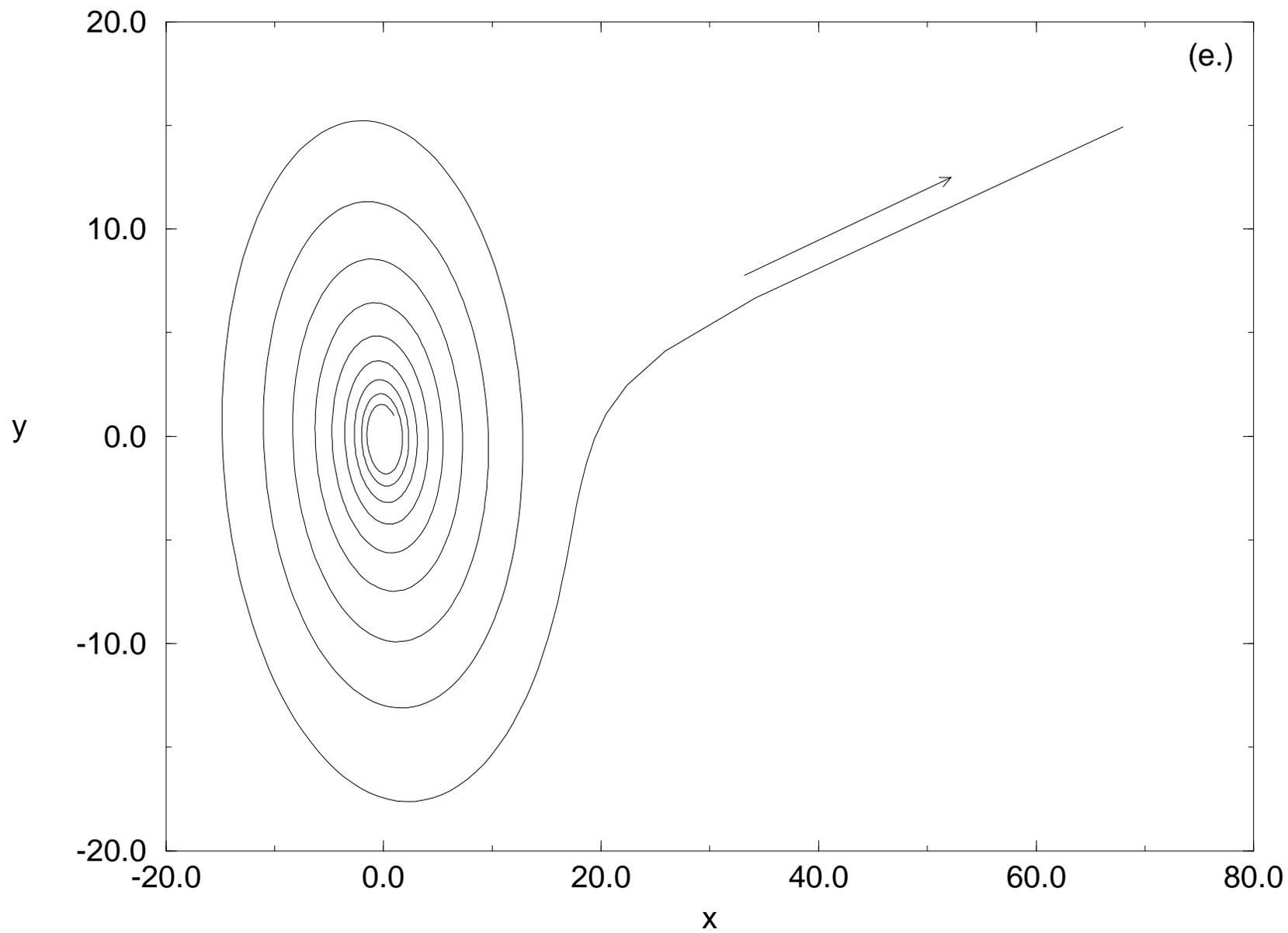

Fig. 2f

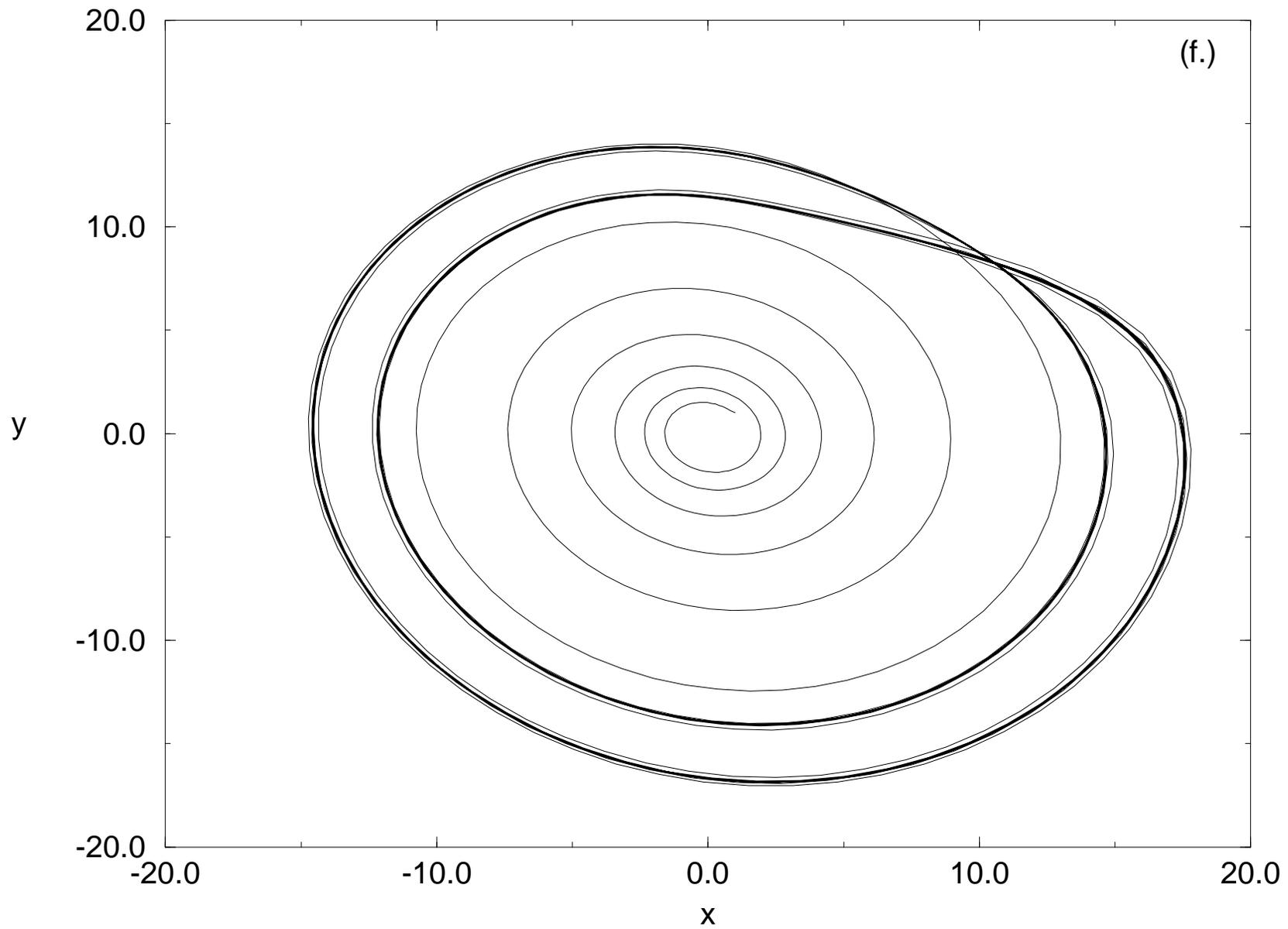

# Fig 3a

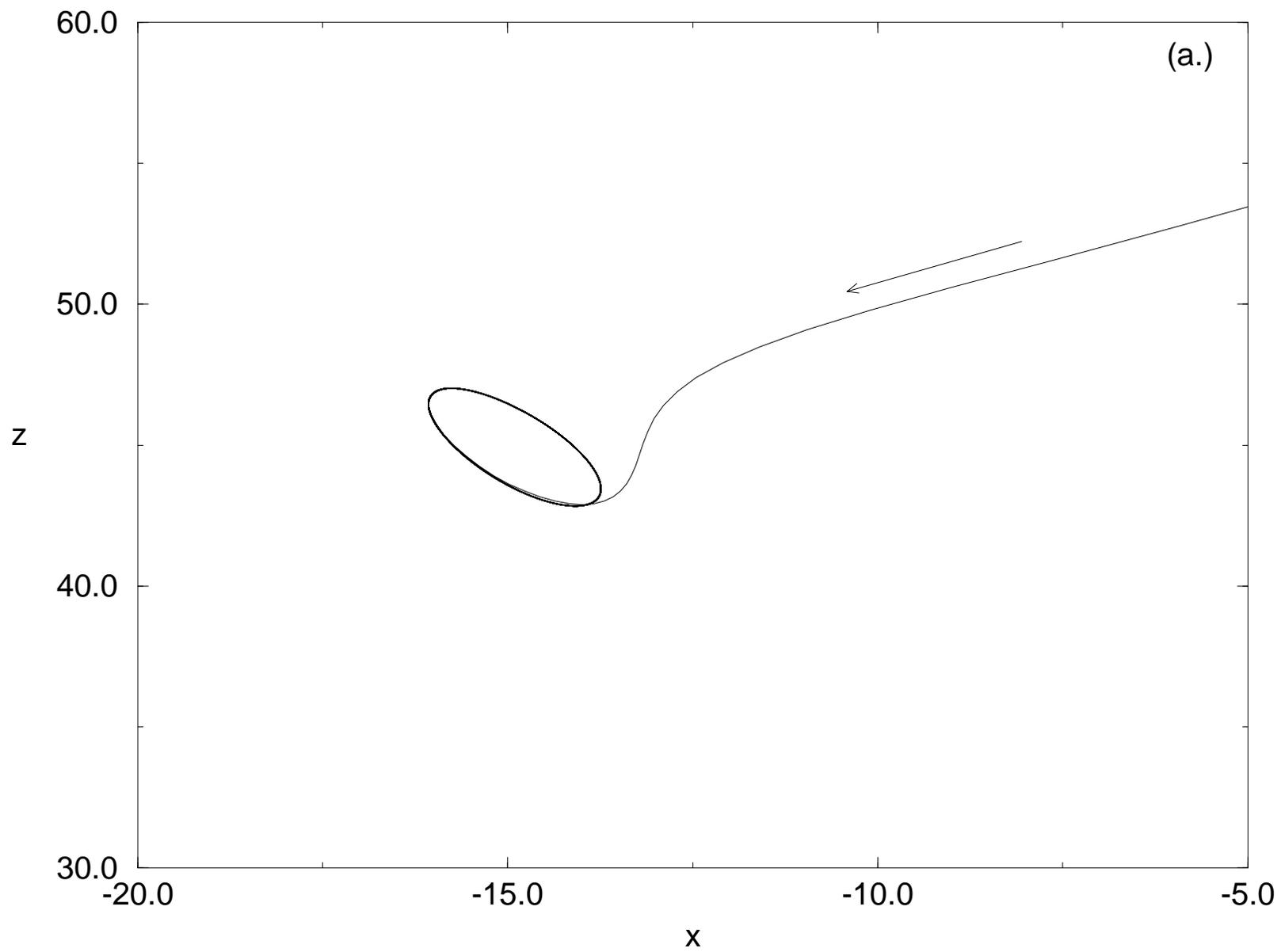

# Fig3c

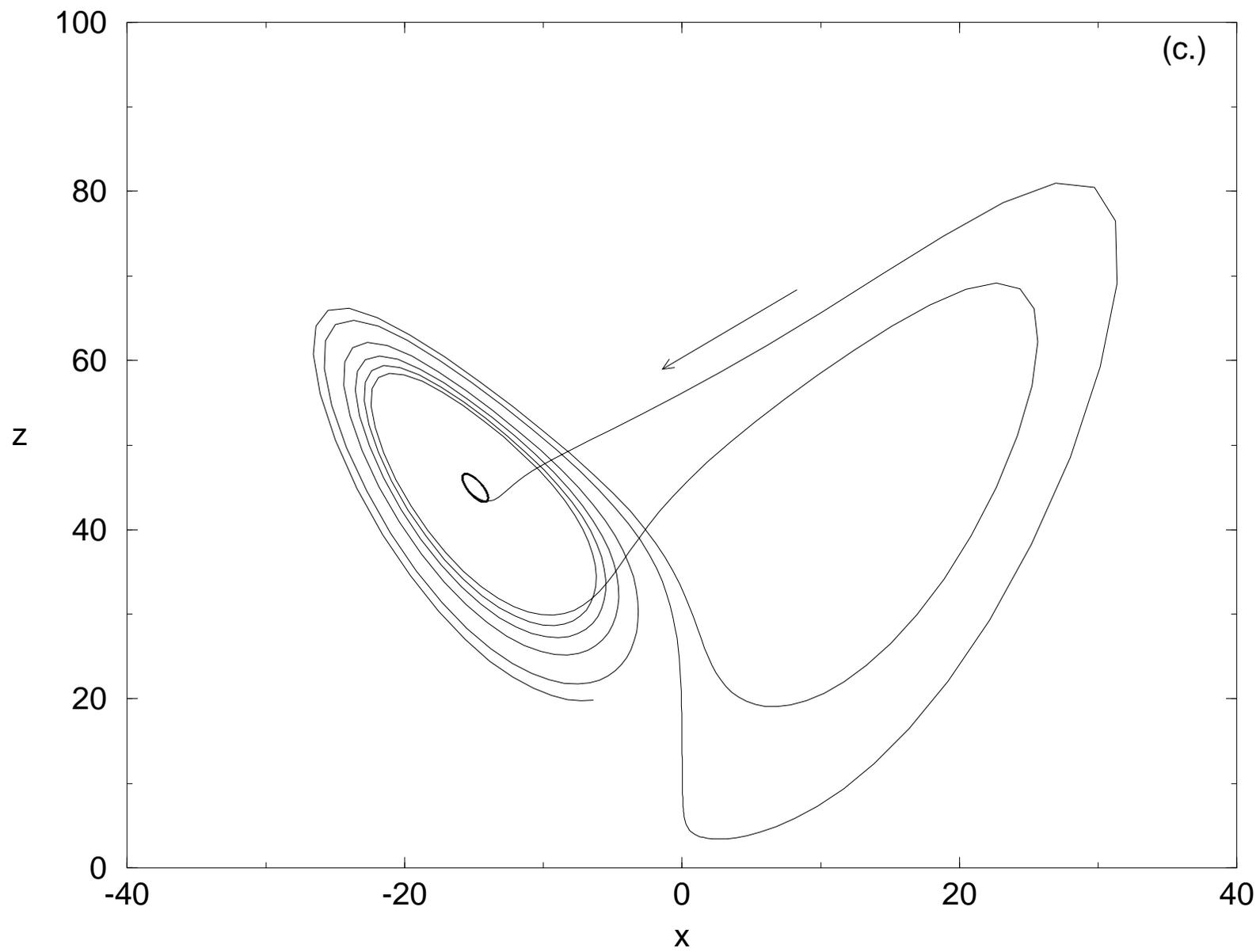